\documentclass[showpacs,preprintnumbers,amsmath,amssymb,12pt]{revtex4}

\usepackage{graphicx}
\usepackage{dcolumn}
\usepackage{bm}
\usepackage{amsmath}
\usepackage{slashed}
\usepackage{mathrsfs}
\usepackage{axodraw4j}
\usepackage{pstricks}
\usepackage{color}
\usepackage{subfigure}
\usepackage{float}
\newcommand{\cc}{c \bar c}
\newcommand{\ups}{\Upsilon}
\newcommand{\jp}{J/\Psi}
\newcommand{\ee}{e^+ e^-}

\begin{document}

\preprint{SDU-ITP202001}

\title{Doubly Heavy Baryon $\Xi_{cc}$ Production in $\Upsilon (1S)$ Decay}
%


\author{Shi-Yuan Li$^{1,*}$, Zhen-Yang Li$^1$, Zong-Guo Si$^1$, Zhong-Juan Yang$^2$,  Xiao Zhang$^1$}
%
 \email{lishy@sdu.edu.cn}
  \affiliation{$^1$ School  of Physics, Shandong University,
Jinan, Shandong, 250100, P.R. China}



\affiliation{$^2$ School of Physics and Technology, University of Jinan,
Jinan, Shandong, 250022, P.R. China}


\thanks{}

\date{\today}

\begin{abstract}
$\Upsilon (1S)$ decay to $\Xi_{cc} +anything$ is studied. It is shown that the corresponding branching ratio can be as significant as  that of $\Upsilon (1S)$ decay to $J/\Psi +anything$.
The non-relativistic heavy quark effective theory framework is employed for the calculation on the decay width.
\end{abstract}

\pacs{Valid PACS appear here}

\maketitle
$\Upsilon (1S)$ \footnote{In the following we use $\Upsilon$ to represent this $1S $ state.}  
 decay is a good arena to study QCD and hadron physics. Several instructive results have been obtained. For example,  recent searches on the exotic XYZ hadrons via the inclusive channel $  \Upsilon \to J/\Psi +anything $   \cite{Shen:2016yzg} and on light tetraquark hadrons  in several channels of $\Upsilon$ decay \cite{Jia:2017zwj} have been made. Both reported negative results.
As a matter of fact, in the energy  region above  $J/\Psi$  mass at BEPC 
and that above $\Upsilon$ mass at B factories, many  exotic XYZ hadrons have been observed (for a recent review, see \cite{Brambilla:2019esw}).  These exotic particles, except those directly couple to the virtual photon in $e^+e^-$ annihilations,    are all produced  from the  {\it decays } of either the exited $c \bar c$ bound states or  the  B hadrons.  On the other hand,   $\Upsilon$  decay is an environment significantly different from those where the exotic particle production is observed.
$\Upsilon$ decays 
 via  the OZI-suppressed ways, i.e., the annihilation of the $b \bar b $ quarks. The  dominant mode ($> 80\%$) is the hadronic one generally refered as '3-gluon' decay \cite{pdg}, and the subsequent hadronization 
  is  a special case of multiproduction.
The negative results \cite{Shen:2016yzg} \cite{Jia:2017zwj} 
 mentioned above can shed light on property of  confinement and the unitarity of the hadronization in multiproduction processes as we have pointed out \cite{Han:2009jw,Jin:2016cpv,Li:2017bqh,Jin:2016vjn}.   
 The experimental facts mentioned above confirm  that the $c \bar c$  pair produced in perturbative process  prefers to transfer into general hadrons like $J/\Psi$ rather than exotic XYZ's in this multiproduction process;
and that for light hadrons, it is also the similar case, i.e., the above  negative experimental results on light exotic hadrons indicate that the dominant decay channels should be    $\Upsilon  \to h's$, with $h's$ referring to mesons as well as baryons.  In one word,
$\Upsilon$ generally decays to  mesons and baryons, 
with  exotic ones 
hardly possible to be observed.
%
%
%
  %
But  the to-date measured decay channels of $\Upsilon$  are much  far from exhausting the total decay width. Especially, almost no
baryon channel is measured 
\cite{pdg}. So measuring the baryon production  is an important task for better understanding the dynamics in
$\Upsilon $ decay.  

Among all the baryons produced in $\Upsilon$ decay, the doubly heavy baryon
  $\Xi_{cc}$  is the most heavy. 
 SELEX and LHCb  have respectively  reported the observations of this kind of baryons  with different mass \cite{Ocherashvili:2004hi,Aaij:2017ueg}.
One of the possibilities can be that  different  SU(2) multi-states of $\Xi_{cc}$ are observed by these two Collaborations. To measure these multi-states, and further to explore SU(3) multi-states,
can surely help to clarify and deepen our knowledge on the property and production mechanism of $\Xi_{cc}$. $\Upsilon$ decay can provide a clean platform
for such measurements.  

There is a further special reason  stands for the  observation on $\Xi_{cc}$ in $\Upsilon$ decay.
   It is noticed that  most of the presented data of $\Upsilon$ decay are upper limits \cite{pdg}. However,   the decay channel $\Upsilon \to J/\Psi +X$ is well measured for several times by
several collaborations  and has attracted wide interests,
which  is important on the study of PQCD and NRQCD (for the full literature list, please see a recent review \cite{Jia:2020csg}).
It was pointed out  that,  based on the  soft $J/\Psi$ spectrum by CLEO measurement which was quite  rough at that time, and on the calculation of the partial width \cite{Li:1999ar},  the dominant contribution could be  $\Upsilon (1S) \to J/\Psi + c \bar c g$. Then the spectrum and branching ratio  is
confirmed by CLEO II \cite{Briere:2004ug,Han:2006vi}
 and later by BELLE \cite{Shen:2016yzg}, though detailed calculations show that several competing sub-processes contribute \cite{He:2019rwt,He:2010cb}.
This fact strongly implies that 
the perturbative  production of  $c \bar c c \bar c$    in $\Upsilon$ decay is significant.
This leads  to that  the double charm   baryon  is hence easily  produced  as argued by the colour connection analysis \cite{Han:2006mpa}. 
 For $c_1\bar c_2 c_3 \bar c_4 g$  system from $\Upsilon$ decay, $c_1\bar c_2$ and $c_3 \bar c_4$ respectively come from a virtual gluon. 
  But $c_1\bar c_4$ and $c_3 \bar c_2$ can respectively be in colour singlet, i.e., the colour space can be reduced as
  $$(3_1 \bigotimes 3^*_4)\bigotimes(3_3\bigotimes 3^*_2)=(1_{14}+8_{14})\bigotimes(1_{23}+8_{23}).$$  This means that such combination of the pair can be colour singlet and easy to translate to  $J/\psi$ for proper invariant mass.  
One  can recognize that the colour space can also  be reduced as
  $$(3_1 \bigotimes 3_3)\bigotimes(3^*_2\bigotimes 3^*_4)=(3^*_{13}+6_{13})\bigotimes(3_{24}+6^*_{24}).$$ In such colour states, the two-charm pair can combine with a light quark to
  become $\Xi_{cc}$ \cite{Han:2006mpa,Jin:2013bra} for proper invariant mass. 
  %
 %
 %
 %
  This simple analysis implies that the production rate of
$\Xi_{cc} + \bar c \bar c g$  is expected  not small once  the $J/\Psi +c\bar cg$ production rate is not small.
 %
 %
%

  In this paper, we devote to study the production of  $\Xi_{cc}$
 in  $\Upsilon$ decay.
  We calculate the corresponding  partial width   and the  momentum distribution of $\Xi_{cc}$.
  Multi-states like $\Xi_{cc}^+$ or $\Xi_{cc}^{++}$ could have different width and lead to quite different feasibility or difficulty in observing  them, but their production mechanism is completely the same in  $\Upsilon$ decay. Therefore  we  do not make any distinction  for the
 investigation on the production.
 In the super B factory, once the center of mass energy is tuned  on the
 $\Upsilon$ resonance,  a large sample of $\Upsilon$ decay data can be obtained and could be employed for the  measurement.  
 The following calculations show that the branching ratio of  $\Xi_{cc}$ production can be order of $10^{-4}$.
   %
   For the $\Upsilon$  decay,
 the process with two charm pairs production  is  easy to be triggered by 3-jet like event shape and  strangeness enhancement (e.g., the $\frac{K}{\pi}$ value) \cite{Jin:2014nva,Han:2006vi}, of which some of the
  the charm meson production events  can be vetoed by lepton pair  or  hadron pair mass around   $J/\Psi$ mass. In this way, one can get a
clean and large sample of events to study the doubly charm baryon multi-states.



   In the process $\Upsilon \to \Xi_{cc}
+ \bar{c} \bar{c} g$,  
   both bottom and the charm quarks are heavy.  
For the initial  bound state, the colour singlet $b \bar b$ pair with  C=-1, it directly leads to the non-relativistic
wave function  formulations \cite{khun,Chang:1979nn,baier,nacht}, where the relative momentum between $b$ and $\bar{b}$
is vanishing, namely  same as the case of positronium. For the final  bound state,
 %
a factorization formulation within the  heavy quark effective theory framework \cite{ms,Li:2017ghe} is employed.
%
  %
One subtle point is that, the non-relativistic formulations are  investigated in the rest frame of each  bound state, respectively; and then a corresponding covariant form of description is obtained, which can be employed in any frame.
Here we start from the initial state:
The differential width of the  process $\Upsilon \to \Xi_{cc}
+ \bar{c} \bar{c} g$
can be formulated as \cite{Li:1999ar}
\begin{equation}
\label{wd}
\frac{d\Gamma}{dR}=\frac{| B_{\Upsilon} < \Xi_{cc}
\bar{c} \bar{c}g| {\cal S} |b\bar{b}(^3S_{1},1)>|^2}{T},
\end{equation}
where $dR$ is the  phase space volume element for
$\Xi_{cc}$ and $ \bar  c$, $\bar{c}, g$ without the constrain of energy momentum conservation;
$\cal S$ is the S-Matrix;
 $B_{\Upsilon}$ is  related to
 the wave function
of  $\Upsilon$ at origin as
\begin{equation}
B_{\Upsilon}=\frac{\Psi_{\Upsilon}(0)}{\sqrt{V}2m_{b}}.
\end{equation}
For convenience, we normalize  all final state particle states to be
$2EV$ (where $E$ is the particle's energy and $V$ is the volume of the total
space).  This normalization is also used for all free quarks in bound states.  For the initial state,  $B_{\Upsilon}$ normalizes the state of  $\Upsilon$  to be
1, so that the width can be directly written as above.
In Eq. (\ref{wd}) the sum over all spin states for  final particles  and average
of the 3 spin states for $\Upsilon$   are not explicitly shown and
 the `time' $T$ is $2\pi\delta(0)$.

For the factorization of the initial bound state,  the width is written, based on the above Equation, as
\begin{equation}
\label{wdp}
d\Gamma= dR' \frac{1}{3} \frac{1}{M^2_{\Upsilon}} |\Psi_{\Upsilon}(0)|^2 | <\Xi_{cc}
\bar{c} \bar{c}g| {\cal T} |b\bar{b}(^3S_{1},1)>|^2.
\end{equation}
Here $dR'=dR (2\pi)^4 \delta^{(4)}(P_i-P_f)$, the factors time T and volume V are cancelled by the $\delta^{(4)}(0)$.
 ${\cal T}$ is the T matrix with $S_{fi}=\delta_{fi}+(2\pi)^4 \delta^{(4)}(P_i-P_f){\cal T}_{fi}$. Sum over all spin states is inexplicitly indicated.

Employing the project operator formulation (e.g., \cite{khun}), and  the radial wave function $R_{\Upsilon}$ to describe the initial bound state, we get the decay amplitude as,
\begin{equation}
 \mathcal{M}_{fi}=\frac{1}{2}\frac{1}{\sqrt{4\pi}}\frac{1}{\sqrt{M_{\Upsilon}}}R_{\Upsilon}(0)Tr[O_{0}({P\mkern-10.5mu/}+M_{\Upsilon})(-{\epsilon\mkern-8.5mu/})].
\end{equation}
$O_{0}$ is the amplitude  for $b \bar b \to \Xi_{cc}\bar{c} \bar{c}g $, with relative momentum of $b\bar b$ vanishing. $P$ and $\epsilon$ are 4-momentum and  polarization vector of $\Upsilon$, respectively.

 In the final state of the $\Upsilon$ decay, the unobserved part X can be divided into a perturbative  part $X_{P}$ and a non-perturbative part $X_{N}$.
 To the lowest-order  (tree level) in PQCD, 
\begin{eqnarray}
\label{amp}
  \mathcal{M}_{fi}&=&\int\frac{d^4q_1}{(2\pi)^4}A_{ij}(k_1,k_2,P_1,P_2,P_3;q_1)\int d^4x_1e^{-iq_1x_1}\nonumber\\
  &\times &<\Xi_{cc}(k)+X_N|\overline{Q_i}(x_1)\overline{Q_j}(0)|0>.
\end{eqnarray}
We assign $k_{1},k_{2},P_{1},P_{2},P_{3},k$ as the momenta of the corresponding  particles,$b, \bar b,  \bar{c}, \bar{c},g, \Xi_{cc} $, respectively, $k_1=k_2 =P/2$.
$A_{ij}(k_1,k_2,P_1,P_2,P_3;q_1)$, which includes the initial wave function,   can be directly read from FIG.1. Both i and j are Dirac and color indices.
 In the matrix element, $X_N$ represents the non perturbative effects.  $Q(x)$ is the Dirac field for charm quark.

\begin{figure}[!h]
  \centering
  \includegraphics[width=0.5\textwidth]{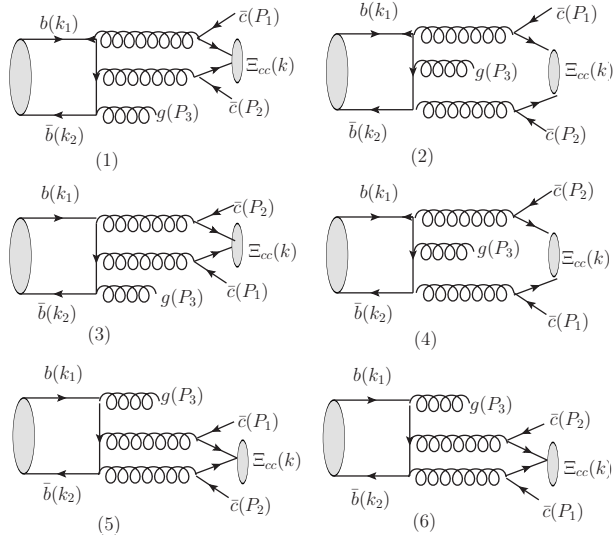}\\
  \caption{Six Feynman diagrams for the 'amplitude' in Eq. (\ref{amp}). The g*  g*  g system are in the same colour, angular momentum and charge conjugation states as those of  $\Upsilon$. The left bubble represents the wave function of $\Upsilon$. $A_{ij}$ does not include the bubble of $\Xi_{cc}$ and the two legs connected to it, which correspond to the matrix element in  Eq. (\ref{amp}).}\label{1}
\end{figure}


   %
Taking the absolute square of the above amplitude, one gets
\begin{eqnarray}
  d\Gamma&=&\frac{1}{2M_{\Upsilon}}\sum_{X_{N}}\frac{d^3k}{(2\pi)^3}\int\frac{d^3P_1}{(2\pi)^32E_1}\frac{d^3P_2}{(2\pi)^32E_2}\frac{d^3P_3}{(2\pi)^32E_3}\nonumber\\
 & \times&(2\pi)^4\delta^4(Q-P_1-P_2-P_3-k)\\
 &\times&\frac{1}{3}\times\frac{1}{3}\times\frac{1}{2}\times\int\frac{d^4q_1}{(2\pi)^4}\frac{d^4q_3}{(2\pi)^4}A_{ij}(k_1,k_2,P_1,P_2,P_3; q_1)\nonumber\\
 &\times&[\gamma^0A^\dag(k_1,k_2,P_1,P_2,P_3;q_3)\gamma^0]_{kl}\int d^4x_1d^4x _3e^{-iq_1x_1+iq_3x_3}\nonumber\\
  &\times&<0|Q_k(0)Q_l(x_3)|\Xi_{cc}+X_N>
  <\Xi_{cc}+X_N|\overline{Q}_i(x_1)\overline{Q}_j(0)|0> \nonumber,
\end{eqnarray}
where the 
spin summation of the baryon $\Xi_{cc}$,   and the polarization and color summation of two anti-charm quarks are implied. Here we take nonrelativistic normalization for the baryon $\Xi_{cc}$. We can eliminate the sum over $X_{N}$ by using translational covariance. Defining the creation operator $a^{\dag}(\mathbf k)$ for $\Xi_{cc}$ with the three momentum $\mathbf k$, we obtain
\begin{eqnarray}
  d\Gamma&=&\frac{1}{2M_{\Upsilon}}\frac{1}{18}\frac{d^3k}{(2\pi)^3}\int\frac{d^3P_1}{(2\pi)^32E_1}\frac{d^3P_2}{(2\pi)^32E_2}\frac{d^3P_3}{(2\pi)^32E_3}\nonumber\\
       &\times&\int\frac{d^4q_1}{(2\pi)^4}\frac{d^4q_3}{(2\pi)^4}A_{ij}(k_1,k_2,P_1,P_2,P_3;q_1)\nonumber\\
        &\times&[\gamma^0A^\dag(k_1,k_2,P_1,P_2,P_3;q_3)\gamma^0]_{kl}\nonumber\\
        &\times&\int d^4x_1d^4x_2d^4x_3e^{-iq_1x_1+iq_3x_3-iq_2x_2}\nonumber\\
        &\times &<0|Q_k(0)Q_l(x_3)a^\dag _{\mathbf k} a_{\mathbf k}\overline{Q}_i(x_1)\overline{Q}_j(x_2)|0>,
\end{eqnarray}
with $q_2=k-q_1$.

\par\setlength\parindent{1em} We use heavy quark effective field theory to deal with the  $\Xi_{cc}$ state. In  $\Xi_{cc}$ rest frame, the heavy quarks move with a small velocity $\upsilon_{c}$.  Hence, the Fourier transformed matrix element can be expanded in $\upsilon_{Q}$ with fields of NRQCD.  The relation between NRQCD fields and Dirac fields Q(x) in the rest frame is
\begin{eqnarray}
   &&Q(x)=e^{-im_c t}\begin{Bmatrix} \psi(x) \\0 \end{Bmatrix}+\mathcal{O}(v_c)+...,
\end{eqnarray}
where $\psi(x)$ is NRQCD field. We will work at the leading order of $\upsilon_{c}$. We denote $\upsilon$ as the velocity of  $\Xi_{cc}$ with $\upsilon^\mu={k^\mu}/{M_{ \Xi_{cc}}}$ to express our result of Fourier transformed matrix element in a covariance way.  Hence, the Fourier transformed matrix element in the rest frame is
\begin{eqnarray}
  &&\upsilon^0\int{d^4q_1}{d^4q_2}{d^4q_3}e^{-iq_1x_1-iq_2x_2+iq_3x_3}\nonumber\\
   &&<0|Q_k(0)Q_l(x_3)a^\dag(\mathbf k)a(\mathbf k)\overline{Q}_i(x_1)\overline{Q}_j(x_2)|0>\nonumber\\
  &=&\int{d^4q_1}{d^4q_2}{d^4q_3}e^{-iq_1x_1-iq_2x_2+iq_3x_3}\nonumber\\
  &&<0|Q_k(0)Q_l(x_3)a^\dag(\mathbf k=0)a^\dag(\mathbf k=0)\overline{Q}_i(x_1)\overline{Q}_j(x_2)|0>. \nonumber\\
\end{eqnarray}
\par\setlength\parindent{1em}Using Eq.(8), the matrix element in Eq.(9) can be expanded with $\psi(x)$ and $\psi^{\dag}(x)$. The spacetime dependence of the matrix element with NRQCD field is controlled by the scale $m_{c}\upsilon_{c}$. At  the leading order of $\upsilon_{c}$ one can neglect the spacetime dependence and the mass of the baryon $M_{\Xi_{cc}}$ is approximated by $2m_{c}$. With the approximation the matrix element in Eq.(9) is
\begin{eqnarray}
 &&<0|\psi{_{\lambda_3}^{a_3}}(0)\psi{_{\lambda_4}^{a_4}}(0)a^\dag a\psi{_{\lambda_1}^{a_1}}(0)\psi{_{\lambda_2}^{a_2}}(0)|0>
\end{eqnarray}
where we suppress the notation  $\mathbf k=0$ and it is always implied that NRQCD matrix elements are defined in the rest frame of $\Xi_{cc}$. The superscripts $a_{i}(i=1,2,3,4)$ are used to label the color of quark fields, while the subscripts $\lambda_{i}(i=1,2,3,4)$ for the quark spin indices. We obtain the matrix element by two parameters, $h_1, h_3$ as following:
\begin{eqnarray}
 &&<0|\psi{_{\lambda_3}^{a_3}}(0)\psi{_{\lambda_4}^{a_4}}(0)a^+a\psi{_{\lambda_1}^{a_1}}(0)\psi{_{\lambda_2}^{a_2}}(0)|0>\nonumber\\
 &=&(\varepsilon)_{{\lambda_4}{\lambda_3}}(\varepsilon)_{{\lambda_2}{\lambda_1}}\cdot(\delta_{{a_1}{a_4}}\delta_{{a_2}{a_3}}+\delta_{{a_1}{a_3}}\delta_{{a_2}{a_4}})\cdot
 h_1\nonumber\\
 &&+(\sigma^n\varepsilon)_{{\lambda_4}{\lambda_3}}(\varepsilon\sigma^n)_{{\lambda_2}{\lambda_1}}\cdot(\delta_{{a_1}{a_4}}\delta_{{a_2}{a_3}}\delta_{{a_1}{a_3}}-\delta_{{a_2}{a_4}})\cdot
 h_3, \nonumber\\
\end{eqnarray}
where $\sigma^i$ ($i=1,2,3$) are Pauli matrices. $\varepsilon=i\sigma^{2}$ is totally anti-symmetric. The parameters  $h_{1}$ and $h_{3}$ are defined as:
\begin{eqnarray}
 h_1&=&\frac{1}{48}<0|[\psi^{a_1}\varepsilon\psi^{a_2}+\psi^{a_2}\varepsilon\psi^{a_1}]a^{\dag}a\psi^{a_2\dag}\varepsilon\psi^{a_1\dag}|0>,\nonumber\\
 h_3&=&\frac{1}{72}<0|[\psi^{a_1}\varepsilon\sigma^n\psi^{a_2}-\psi^{a_2}\varepsilon\sigma^n\psi^{a_1}]a^{\dag}a\psi^{a_2\dag}\sigma^n\varepsilon\psi^{a_1\dag}|0>.
 \nonumber\\
\end{eqnarray}
$h_{1}(h_{3})$ represents the probability for a cc pair in a $^{1}{S}_{0}(^{3}{S}_{1})$ state and in the color state of 6($\bar{3*}$) to transform into the baryon.
It is the Pauli exclusion principle determines that only these two kinds of combination of colour and spin states, which are asymmetric,  are possible \cite{ms}.
With these results the Fourier transformed matrix element in Eq.(9) can be expressed as:
\begin{eqnarray}
&&\upsilon^0\int{d^4x_1}{d^4x_2}{d^4x_3}e^{-iq_1x_1-iq_2x_2+iq_3x_3}\nonumber\\
&&<0|Q{{_k}^{a_3}}(0)Q{{_l}^{a_4}}(x_3)a^\dag(\mathbf k)a(\mathbf k)\overline{Q}{{_i}^{a_1}}(x_1)\overline{Q}{{_j}^{a_2}}(x_2)|0>\nonumber\\
&=&(2\pi)^4\delta^4(q_1-m_c\upsilon)(2\pi)^4\delta^4(q_2-m_c\upsilon)(2\pi)^4\nonumber\\
&&\delta^4(q_3-m_c\upsilon)\times[-(\delta_{a_1a_4}\delta_{a_2a_3}+\delta_{a_1a_3}\delta_{a_2a_4})\nonumber\\
&&(\widetilde{P}_{v}C\gamma_{5}P_{v})_{ji}(P_{v}\gamma_{5}C\widetilde{P}_{v})_{lk}h_{1}\nonumber\\
          &&+(\delta_{a_1a_4}\delta_{a_2a_3}-\delta_{a_1a_3}\delta_{a_2a_4})(\widetilde{P}_{v}C\gamma^{\mu}P_{v})_{ji}\nonumber\\
          &&(P_{v}\gamma^{\nu}C\widetilde{P}_{v})_{lk}(\upsilon_{\mu}\upsilon_{\nu}-g_{\mu\nu})h_{3}]+...
\end{eqnarray}
where $P_{\upsilon}=\frac{1+\gamma\cdot\upsilon}{2}$,$\widetilde{P}_{\upsilon}=\frac{1+\widetilde{\gamma}\cdot\upsilon}{2}$; $C=i \gamma^2 \gamma^0$, the charge conjugation operator.\\

  With the above formula, we obtain the decay width as following:
\begin{eqnarray}
  d\Gamma&=&\frac{16 \pi^4\alpha^{5}_{s}|R_{\Upsilon}(0)|^{2}M_{\Xi_{cc}}}{9M^{2}_{\Upsilon}}\frac{1}{[{(P_{1}+k/2)^{2}(P_{2}+k/2)^{2}}]^{2}}\nonumber\\
  &&\sum^8_{c=1}(\sum^{6}_{\xi=1}\overline{A}^{abc}_{\xi})(\sum^{6}_{\zeta=1}\overline{A}_{\zeta}^{*a'b'c})H^{aba'b'}\\
 &&\frac{d^{3}k}{(2\pi)^{3}2E_{k}}\prod^{3}_{i=1}\frac{d^{3}P_{i}}{(2\pi)^{3}2E_{i}}(2\pi)^4\delta^4(Q-P_1-P_2-P_3-k)\nonumber,
\end{eqnarray}
where
\begin{eqnarray}
  H^{aba'b'}&=&-(Tr[T^{a}T^{a'}T^{b}T^{b'}]+Tr[T^{a}T^{a'}]Tr[T^{b}T^{b'}])\times h_{1}\times B_{1}\nonumber\\
     &&+(Tr[T^{a}T^{a'}T^{b}T^{b'}]-Tr[T^{a}T^{a'}]Tr[T^{b}T^{b'}])\times h_{3}\times B_{2},\nonumber\\
\end{eqnarray}
\begin{eqnarray}
B_{1}&=&Tr[\gamma^{\alpha}({P_{2}\mkern-15.5mu/}-m_{c})\gamma^{\alpha'}P_{v}\gamma_{5}\widetilde{P}_{v}\gamma^{\beta'}(-{P_{1}\mkern-15.5mu/}
-m_{c})\gamma^{\beta}\widetilde{P}_{v}\gamma_{5}P_{v}],\nonumber\\
\end{eqnarray}
\begin{eqnarray}
B_{2}&=&Tr[\gamma^{\alpha}({P_{2}\mkern-15.5mu/}-m_{c})\gamma^{\alpha'}P_{v}\gamma_{\nu}\widetilde{P}_{v}\gamma^{\beta'}
(-{P_{1}\mkern-15.5mu/}-m_{c})\gamma^{\beta}\widetilde{P}_{v}\gamma_{\rho}P_{v}](\upsilon_{\rho}\upsilon_{\nu}-g_{\mu\nu}).
\end{eqnarray}
The function $\overline{A}_{\xi}(\xi=1,2,3,4,5,6)$ are given in Appendix A.  

The radial wave function for $\Upsilon$ at origin can be  obtained, e.g., by fitting its leptonic decay width.
%
  On the other hand, 
    the
   value of $h_1$ and $h_3$, is difficult to  be obtained.  There are  no experiment results now. Here  we employ a potential model with  the radial wave function $R_{cc}(r)$ at origion \cite{Berezhnoy:1998aa} to get the numerical value of $h_3$
\begin{equation}
h_3=\frac{|R_{cc}(0)|^2}{4\pi},
\end{equation}
with its value to be $0.0287 GeV^3$.  There is no practical model for
$h_1$, which can be  taken as a free parameter, the reason  is explained later.
In the numerical calculations, we take $\Psi_{\Upsilon}(0) = 2.194GeV^{3/2}$, 
 $M_{\Upsilon}=9.46GeV, M_{\Xi}=3.621GeV, m_{b}=4.73GeV, \alpha_{s}(m_c)=0.253 $.   $m_{c}/m_b$   is taken to be   parameter,  and  
 the dependence  of branching ratio on $m_{c}/m_{b}$ is studied as shown in FIG.2.

  With  $m_{c}/m_b=0.25$ 
  the partial  width is $\Gamma=(0.0126h_1+0.240h_3)$ KeV.  Here we see that the perturbative part corresponding to $h_1$ is much smaller than that of $h_3$. So if there is no special enhancement on  $h_1$, this part of contribution can not be significant. Here for simplicity we
   take $h_{1}=h_{3}$, and  the decay width is 7.256eV, leading to  the branching ratio as $1.34\times10^{-4}$. The $\Xi_{cc}$ momentum distributions are shown in FIG.3 and FIG. 4. The momentum distributions of $\bar{c}$ are shown in FIG.5 and FIG 6.

  \begin{figure}[!h]
  \centering
  \includegraphics[width=0.4\textwidth]{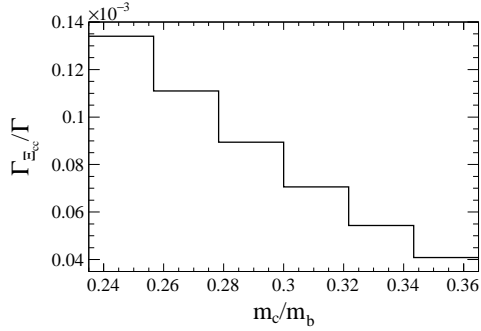}\\
  \caption{Dependence of branching ratio on $m_{c}/m_{b}$.}\label{1}
\end{figure}

\begin{figure}[!h]
  \centering
  \includegraphics[width=0.4\textwidth]{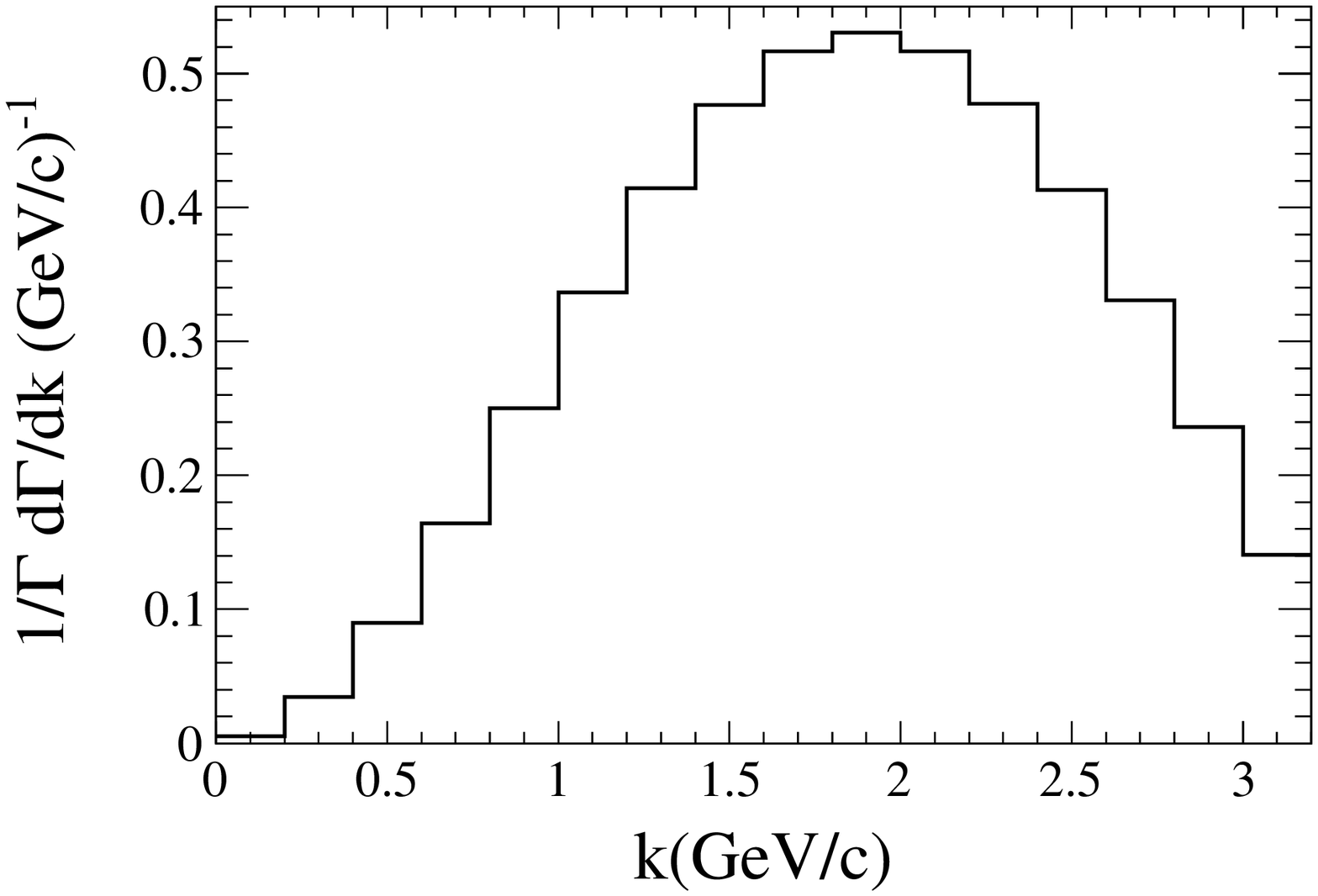}\\
  \caption{The momentum distribution of $\Xi_{cc}$,$h_3=0$}\label{1}
\end{figure}
\begin{figure}[!h]
  \centering
  \includegraphics[width=0.4\textwidth]{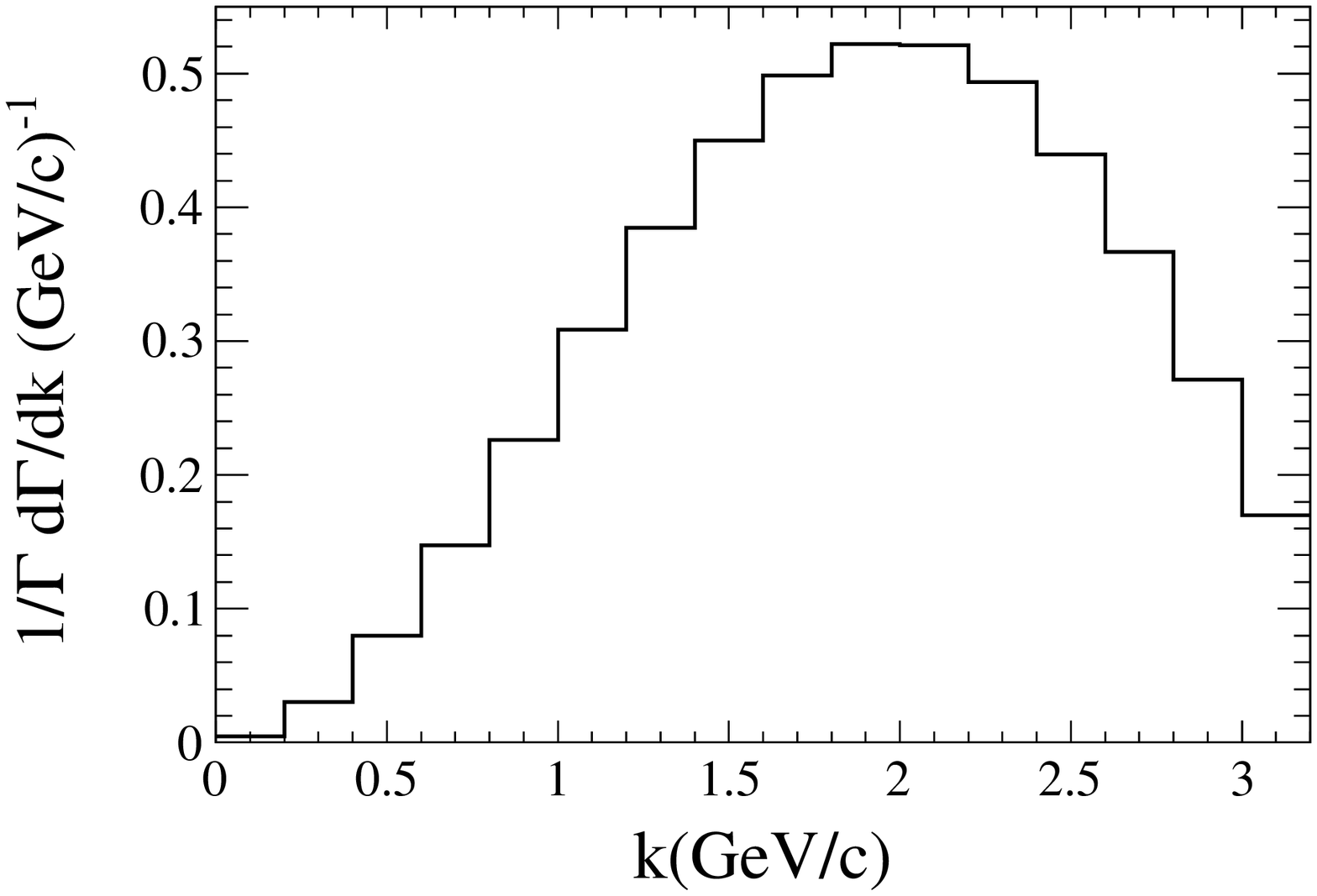}\\
  \caption{The momentum distribution of $\Xi_{cc},h_1=0$}\label{1}
\end{figure}
\begin{figure}[!h]
  \centering
  \includegraphics[width=0.4\textwidth]{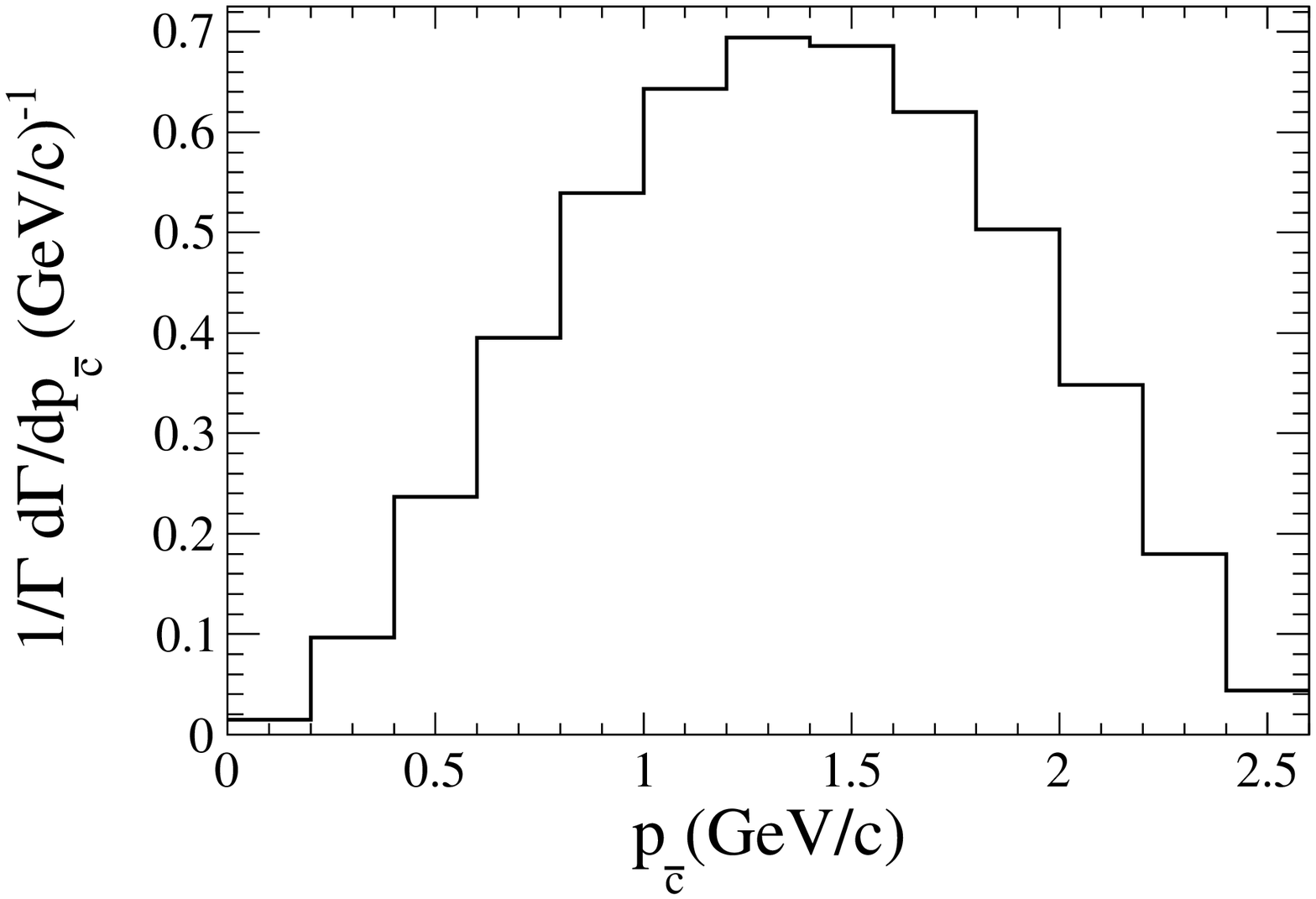}\\
  \caption{The momentum distribution of $\bar{c},h_3=0$}\label{1}
\end{figure}
\begin{figure}[!h]
  \centering
  \includegraphics[width=0.4\textwidth]{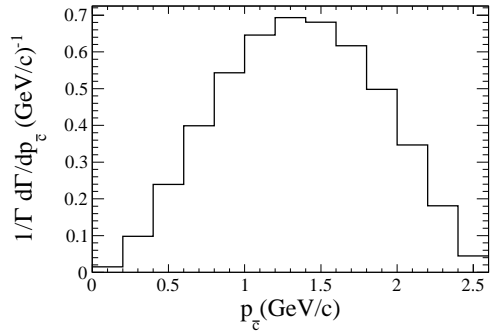}\\
  \caption{The momentum distribution of $\bar{c},h_1=0$}\label{1}
\end{figure}
The experiment of BELLE in 2016 has collected $102\times10^6$ $\Upsilon$ events \cite{Shen:2016yzg,Jia:2020csg}. So it is possible to  make a scan on the $\Xi_{cc}$ production. In the future,    further precise measurement on the production of $\Xi_{cc}$ can  even be made with more large luminosity at BELLE2. Similar  productions characteristic of the partonic   state with four charm (anti)quarks can also be studied in $\Upsilon$ decay.

\section*{Acknowledgments}
 This work is supported by National Natural Science Foundation of China (grant Nos. 11635009, 11775130) and the Natural Science Foundation of Shandong Province (grant ZR2017MA002).

\appendix
\section{}
The functions $\bar A_\xi(\xi=1,...,6)$ in the decay width are:
\begin{eqnarray}
\bar{A_1} &  = &Tr^s[T^cT^bT^a]\frac{1}{[(q-P_3)^2-m^2][(q-P_1-k/2)^2-m^2]}\nonumber\\
&\times & Tr[\slashed{\varepsilon}^*(P_3)(m+\slashed{P_3}-\slashed{q})\gamma_{\alpha}(\slashed{q}-\slashed{P_1}-\frac{\slashed k}{2}+m)
\gamma_{\beta}(M+\slashed{P})\slashed{\epsilon}]\nonumber\\
\bar{A_2}&=&Tr^s[T^bT^cT^a]\frac{1}{[(P_2+k/2-q)^2-m^2][(q-P_1-k/2)^2-m^2]}\nonumber\\
&\times & Tr[\gamma_{\alpha}(m+\slashed{P_2}+\frac{\slashed k}{2}-\slashed{q})\slashed{\varepsilon}^*(P_3)(\slashed{q}-\slashed{P_1}-\frac{\slashed k}{2}+m)
\gamma_{\beta}(M+\slashed{P})\slashed{\epsilon}]\nonumber\\
\bar{A_3}&=&Tr^s[T^cT^aT^b]\frac{1}{[(P_3-q)^2-m^2][(q-P_2-k/2)^2-m^2]}\nonumber\\
&\times& Tr[\slashed{\varepsilon}^*(P_3)(m+\slashed{P_3}-\slashed{q})\gamma_{\beta}(\slashed{q}-\slashed{P_2}-\frac{\slashed k}{2}+m)
\gamma_{\alpha}(M+\slashed{P})\slashed{\epsilon}]\nonumber\\
\bar{A_4}&=&Tr^s[T^aT^cT^b]\frac{1}{[(P_1+k/2-q)^2-m^2][(q-P_2-k/2)^2-m^2]}\nonumber\\
&\times &Tr[\gamma_{\beta}(m+\slashed{P_1}+\frac{\slashed k}{2}-\slashed{q})\slashed{\varepsilon}^*(P_3)(\slashed{q}-\slashed{P_2}-\frac{\slashed k}{2}+m)\gamma_{\alpha}(M+\slashed{P})\slashed{\epsilon}]\nonumber\\
\bar{A_5}&=&Tr^s[T^bT^aT^c]\frac{1}{[(P_2+k/2-q)^2-m^2][(q-P_3)^2-m^2]}\nonumber\\
&\times& Tr[\gamma_{\alpha}(m+\slashed{P_2}+\frac{\slashed k}{2}-\slashed{q})\gamma_{\beta}(\slashed{q}-\slashed{P_3}+m)\slashed{\varepsilon}^*(P_3)(M+\slashed{P})\slashed{\epsilon}]\nonumber\\
\bar{A_6}&=&Tr^s[T^aT^bT^c]\frac{1}{[(P_1+k/2-q)^2-m^2][(q-P_3)^2-m^2]}\nonumber\\
&\times & Tr[\gamma_{\beta}(m+\slashed{P_1}+\frac{\slashed k}{2}-\slashed{q})\gamma_{\alpha}(\slashed{q}-\slashed{P_3}+m)\slashed{\varepsilon}^*(P_3)(M+\slashed{P})\slashed{\epsilon}]
\end{eqnarray} 
Here $Tr^s[...]$ means only keeping the symmetric part; $m=m_b, M=M_{\Upsilon},q=P/2$. $\varepsilon(P_3)$ is the polarization vector of the gluon with momentum $P_3$.

\end{document}